\documentclass[useAMS,usenatbib]{mn2e}

\usepackage[pdftex]{graphicx}

\usepackage{relsize}

\newcommand{\hi}{%
  \relax
  \ifmmode
    \textrm{\textsc{HI}}
  \else
    \textsc{H{\smaller}I}
  \fi
}

\newcommand{\simgt}{\lower.5ex\hbox{$\; \buildrel > \over \sim \;$}}
\newcommand{\simlt}{\lower.5ex\hbox{$\; \buildrel < \over \sim \;$}}

\title[The Shape of Dark Matter Haloes IV. The Structure of Stellar Discs in Edge-on Galaxies]{The Shape of Dark Matter Haloes\\
IV. The Structure of Stellar Discs in Edge-on Galaxies}
\author[S. P. C. Peters et al.]{S. P. C. Peters$^{1}$,
G. de Geyter${^2}$,
P. C. van der Kruit$^{1}$\thanks{For more information, please contact P.C. van der Kruit at vdkruit@astro.rug.nl.} 
and K. C. Freeman$^{3}$\\
$^{1}$Kapteyn Astronomical Institute, University of Groningen, P.O.Box 800, 9700AV Groningen, the Netherlands\\
$^{2}$Sterrenkundig Observatorium, Universiteit Gent, Krijgslaan 281-S9, 9000 Gent, Belgium\\
$^{3}$Research School of Astronomy and Astrophysics The Australian National University, Cotter Road Weston Creek, ACT 2611,\\
Australia}

\begin{document}

\date{Accepted 2015 month xx. Received 2015 Month xx; in original form 2015 Month xx}
\pagerange{\pageref{firstpage}--\pageref{lastpage}} \pubyear{2015}

\maketitle

\label{firstpage}

\begin{abstract}
We present optical and near-infrared archival observations 
of eight edge-on galaxies.  
                    These observations are used to model 
the stellar content of each galaxy using the FitSKIRT software package.
                    Using FitSKIRT, we can self-consistently 
model a galaxy in each band simultaneously while treating for dust.
                    This allows us to accurately measure both 
the scale length and scale height of the stellar disc, plus
                     the shape parameters of the bulge.
                    By combining this data with the previously 
reported integrated magnitudes of each galaxy, 
                     we can infer their true luminosities.
					We have successfully 
modelled seven out of the eight galaxies in our sample.
					We find that stellar 
discs can be modelled correctly, but have not been able to model 
the stellar bulge reliably. 
					Our sample consists for 
the most part of slow rotating galaxies, and we find that the 
average dust layer is much thicker than what is reported for 
faster rotating galaxies.
\end{abstract}


\section{Introduction}
This is the fourth paper in a series of five where we try to constrain the 
flattening of dark matter haloes from observations of the thickness and
velocity dispersion of the \hi layer in edge-on late-type dwarf galaxies.
For this analysis we need to correct for the influence of the stellar disks
on the dynamics of the gas and therefore we provide in this paper fits
to the light distribution in the stellar disks of our sample galaxies.

Edge-on galaxies offer a unique perspective on the distribution of stars and 
dust in galaxies. A major advantage of the edge-on view is the ability to 
resolve the vertical distribution of both the stars and the dust, and this is
exactly the reason that in this series of papers we study edge-ons.
A well-known feature of edge-on spiral galaxies are truncations at the 
outer edge of their stellar discs, as discovered in edge-on galaxies back in 
1979 \citep{vdk79}.
Subsequent authors have confirmed the presence of truncations, such as the 
study of 34 edge-on galaxies by \citet{Kregel2002}, who found that at least 
20 of these galaxies ($\sim\!60\%$ of the sample) have truncations.
Not all galaxies are truncated. Some even appear to have an upturn in
their radial profiles \citep{Erwin2008A}, although \citet{kf11} noted
that many of these, e.g. NGC\,3310, show signs of merging or other
distortions in the outer parts. Some galaxies extend out to many 
scalelengths, such as NGC\'300 \citep{Joss2005}. In this
series op papers we do not concern 
ourselves with the outer features, as we model the \hi only in the inner 
parts in order to set limits of the three-dimensional shape of dark matter 
halos. In Paper V we determine out to which radius we feel our data justify 
fitting for the  shape of the dark matter halos; the actual radii 
adopted for analysis vary from less than 1 out to typically 4 or 5 
scalelengths, so we do not go out into these areas of extended stellar disks
or upturns. Also, if any of our systems would have a truncation, our analysis
is restricted to the areas within that truncation.

A major drawback to the edge-on perspective is the more complex 
geometry, which implies that each position $x$ along the major 
axis is a superposition of light emitted at a range of radii $R$. 
Various authors have therefore set out to disentangle edge-on 
galaxies and extract their radial structure.
One approach to this is to deproject the edge-on image using 
the inverse Abel transform \citep{Binney1987A}.
This method was first applied by \citet{Florido2001A,Florido2006A} 
in a study of the mid-plane of edge-on 
galaxies in the near infrared.
\citet{Pohlen2007} extended the method to study both the 
radial and vertical distribution of eleven edge-on galaxies.
UGC\,7321 was studied in this way by \citet{OBrien2010D}.

Dust can seriously hamper this type of deprojection, as 
there is no simple way to incorporate its complex interplay of 
scattering, absorption and emission, in the inverse Abel transform. 
An alternative strategy for deriving the radial properties 
of the galaxy is to model the entire galaxy including the dust.
By using appropriate fitting algorithms, the model can be 
tweaked in various ways until it matches the observations. 
Various authors applied this method to edge-on galaxies, 
such as \citet{Xilouris1999} and \citet{Bianchi2007}. 

\textsc{FitSKIRT} was developed by \citet{Geyter2013A} as 
an automatic fitting extension for the \textsc{SKIRT} 3D-continuum 
Monte-Carlo radiative transfer code 
\citep{Baes2001A,Baes2001B,Baes2003A,Baes2011A}.
\textsc{FitSKIRT} makes use of the \textsc{GAlib} genetic 
algorithm to fit a parameterised model of a galaxy \citep{galib}.
In the so-called oligochromatic mode, the code can fit multiple 
bands simultaneously.
This mode was used by \citet{deLooze2012B} to model edge-on 
galaxy NGC\,4565 from the UV to the mid-infrared  
in a self-consistent manner.
This method has been applied by \citet{deGeyter2014} to a set 
of twelve edge-on galaxies selected from the CALIFA survey.

In the previous paper (number III in thsi series), 
we have modelled the \hi structure and 
kinematics of eight edge-on dwarf galaxies, as part of our project 
to model their dark matter halo by inferring its properties from 
the baryonic content.
In the current paper, we continue this project by modelling the 
stellar and dust components of these eight edge-on galaxies.
We have collected a large set of observations, which we present in 
Section \ref{sec:stellardata}.
The models and strategy by which we fit the data are presented in 
Section \ref{section:stellardatafitting}.
The fits to the data using \textsc{FitSKIRT} are presented in 
Section \ref{sec:stellarresults}, after which we discuss the 
results in Section \ref{sec:stellardiscussion} and conclude 
the paper in Section \ref{sec:stellarconclusions}.

\section{Data}\label{sec:stellardata}
\textsc{FitSKIRT} can model observations ranging from the UV to the infrared.
The more bands are available to it, the better the models can be constrained.
In this papers, we continue the analysis of the eight edge-on 
galaxies from the previous papers in this series.
See Table 2 in Paper I for an overview of the sample.
In order to get the largest dataset, a wide range of archives, 
both online and off-line, has been explored.
For most galaxies, a large set of archived observations were found.
Unfortunately, the signal-to-noise ratio in many of these (survey) 
observations was insufficient, as the low surface brightness nature 
of most of these galaxies necessitated longer exposure times than were used.
In total, we use data from ten telescopes in nine bands, giving a 
minimum of three bands per galaxy, and a maximum of nine.
A full overview of the observations that we used is shown in Table 
\ref{tbl:stellarobservations} in the online Appendix.

\subsection{Data Reduction}
\textsc{FitSKIRT} does not require the data to be calibrated, so we 
skipped that step in the reduction.
As our observations come from a wide range of telescopes, each required its 
own, appropriate treatment.

\begin{description}
\item[ {\bf 2.3m ANU Advanced Technology Telescope} ] \hfill \\
The 2.3m ANU Advanced Technology Telescope
with the CASPIR near-IR camera, 
located at Siding Spring observatory in Australia, 
was used to observe several of the galaxies in the Kn- and H-bands.
The data were archived on tape and was originally 
intended for use in \citet{OBrien2010D}.
The CASPIR instrument has dedicated software for reducing 
the data, which suffered from code-rot and did not work correctly any more. 
With support from Peter McGrergor at the Research School of 
Astronomy \& Astrophysics in Canberra, Australia, along with 
the manual and source code for CASPIR, we managed to reconstruct 
the required workflow and re-implement this in \textsc{Python}.

During an observing run, various types of observations are being made: 
bias, dark, flat, sky and object.
First, we create the combined bias and this from all other types of frames.
CASPIR has a non-linear response and it was thus required to linearise the 
flat, dark sky and object frames.
The dark currents were then removed from the sky, object and flat frames, 
after which the sky and object frames were flat-fielded.
The sky in the near infrared is bright and varies rapidly.
The main observations were therefore taken as one short object frame 
followed by a sky frame.
The sky frames before and after each object frame were used to subtract 
the sky from a particular object frame.
The field-of-view of CASPIR only covers a fraction of a galaxy, 
so a dithering pattern was used to cover the entire galaxy.
The information on the exact dithering pattern was unavailable and 
we thus resorted to manually calibrating the coordinate system 
of each frame, using \textsc{iraf}.
Each galaxy consisted of more than 50 object frames.
Afterwards we use the \textsc{Montage} tool-kit\footnote{Available at 
montage.ipac.caltech.edu/.} to automatically create a montage of all 
these frames.
This process automatically performs background rectification between 
the various frames.

\item[ {\bf 3.9m Anglo-Australian Telescope} ] \hfill \\ 
The IRIS2 instrument on the 3.9m Anglo-Australian Telescope, 
located at Siding Spring observatory in Australia, was used for 
various Ks- and H-band observations.
Observations were available on tape and were originally intended for 
use in \citet{OBrien2010D}.
In contrast to the  2.3m ANU Advanced Technology Telescope observations, 
the field-of-view is much larger and covers the entire galaxy and the 
surrounding patch of sky.
The IRIS2 instrument has custom software, which automatically runs 
through the entire workflow and returns the reduced science-ready frame.

\item[ {\bf ANU 40-inch Telescope}      ] \hfill \\
The Australian National University (ANU) 40-inch telescope, 
located at Siding Spring Observatory in Australia, was used 
to observe several of our galaxies, as intended for use in 
\citet{OBrien2010D}.
The data was archived on tapes. 
The traditional  \textsc{IRAF} workflow of bias and flat-fielding 
removal was performed to reduce the data. 

\item[ {\bf CTIO 0.9-meter Telescope}  ] \hfill \\
The R-band image for ESO115-G021 was taken using the 
0.9-meter Telescope located at the Cerro Tololo Inter-American 
Observatory (CTIO) in Chili.
The observation was taken as part of the Spitzer Local Volume 
Legacy project and was provided science-ready online at the 
NASA/IPAC Extragalactic Database (NED)\footnote{Available at 
ned.ipac.caltech.edu/.}.

\item[ {\bf Danish 1.54-meter Telescope}              ] \hfill \\
The Danish 1.54-m Telescope is located at the 
European Southern Observatory (ESO) La Silla site in Chili.
The R-band image for ESO\,274-G001 was previously published by 
\citet{Rossa2003A} and was available science-ready online via NED.

\item[ {\bf ESO\,1-meter Schmidt Telescope}            ] \hfill \\
The ESO-LV survey \citet{Lauberts1988A} digitised 606 blue and 606 
red ESO\,photographic survey plates.
The data was available science-ready through NED.

\item[ {\bf ESO\, La Silla Schmidt telescope}             ] \hfill \\
The ESO\,red-band survey has been digitised using the MAMA microdensitometer 
and provided as science-ready data and available through a 
virtual-observatory (VO) interface. These images were 
originally part of the ESO(B) Atlas, taken with the 
ESO\,1-m Schmidt telescope at La Silla, Chile \citep{Lauberts1982}.

\item[ {\bf Palomar 48-inch Schmidt Telescope}   ] \hfill \\
The 48-inch Schmidt telescope at Palomar Observatory 
in the United States  was used to create the Palomar Sky Survey (POSS). 
This survey was later digitised into the Digitized Sky Survey (DSS).
The data was provided science-ready through a virtual-observatory interface.

\item[ {\bf Spitzer Space Telescope IRAC} ] \hfill \\
The Spitzer Space Telescope archive had
 $3.6\,\mu$m observations of all our galaxies and for 
several also  $4.5\,\mu$m observations.
The observations were taken for a range of projects 
\citep{Dale2009A, Sorce2012A, Engelbracht2008A, Sheth2010A, Radburn-Smith2011A, MacLachlan2011A}.
The Spitzer Heritage Archive\footnote{Available at 
sha.ipac.caltech.edu/applications/Spitzer/SHA.} 
provided direct online access to the science-ready frames.

\item[ {\bf UK 48-inch Schmidt Telescope}        ] \hfill \\
The UK 48-inch Schmidt, located at Siding Spring in 
Australia, was used in an all-sky survey. 
This was digitised into the DSS and was available 
science-ready through a virtual-observatory interface.
\end{description}

\subsection{Astrometric Calibration}
The astrometric solution associated with the various observations would 
often be in disagreement between bands, causing galaxies to be offset 
slightly between bands. 
We therefore used the \textsc{solve-field} program, which is part of the 
astrometry.net project, to fit a new astrometric calibration to all the 
observations.
The telescopes and cameras used in this study are of a large variety
and consequnetly not a single limiting magnitudes in the bands can be 
given for our sample. In general, our photometry does not go as deep as the
\citet{Comeron2011B} study of edge-on galaxies, but our interest is in the 
brighter levels in order to constrain the light distribrutions for our
modelling.

\subsection{Masking}
To avoid flux contamination, it was required to mask all fore- and 
background objects. 
For each galaxy, the band with the most prominent stars was selected.
We used the \textsc{ds9} display tool to draw regions on each object, 
after which a custom \textsc{Python} program set the pixels in these 
regions to a value of zero.
Pixels with a value of zero are ignored by \textsc{FitSKIRT}, and we 
thus do not need to interpolate over them.
We also used this program to draw a polygon around each galaxy, beyond 
which the image was masked. 
In most cases there was only little masking required. 
The galaxies closer to the galactic plane, however, 
required far more extensive masking, with the total 
amount of masks drawn for ESO\,274-G001 well beyond a thousand.
The region masks were copied to the other bands using \textsc{ds9}.
The files were then inspected and additional masking applied where required.

\subsection{Normalization}
To avoid overly long computations, all images were rotated to align 
their major axis with the horizontal image axis, and then shrunk 
down to a width of $\sim700$ pixels. 
We subtract the average background from the image.
The image was then divided by the total flux, such that the sum 
of the image became one.
This last step is required by \textsc{FitSKIRT}, as it prevents 
a particularly bright band from dominating the fitting result.

\begin{figure*}
\centering
\resizebox{0.48\textwidth}{!}{\includegraphics{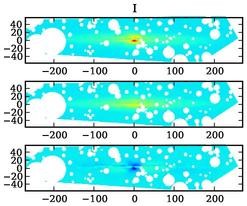}}
\resizebox{0.48\textwidth}{!}{\includegraphics{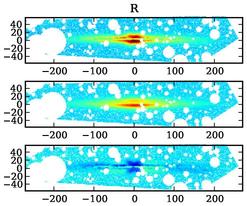}}
\resizebox{0.48\textwidth}{!}{\includegraphics{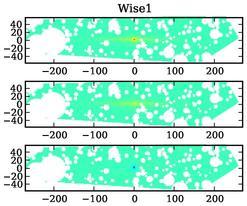}}
\caption[IC\,2531 - stellar fit]{Stellar decompositions for IC\,2531. Each band consists of three panels. Top panels show the observation. Middle panels show the best-fit models. The lower panels show the difference maps. Colour scaling is equal between the three panels. The scale of the images is in arcsec.}
\end{figure*}

\begin{figure*}
\centering
\resizebox{0.48\textwidth}{!}{\includegraphics{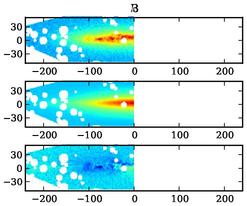}}
\resizebox{0.48\textwidth}{!}{\includegraphics{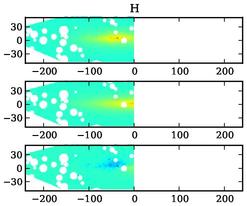}}
\resizebox{0.48\textwidth}{!}{\includegraphics{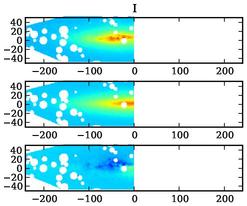}}
\resizebox{0.48\textwidth}{!}{\includegraphics{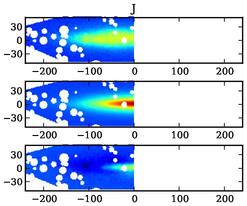}}
\resizebox{0.48\textwidth}{!}{\includegraphics{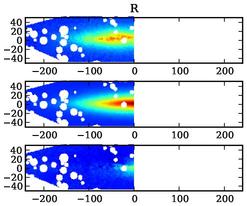}}
\resizebox{0.48\textwidth}{!}{\includegraphics{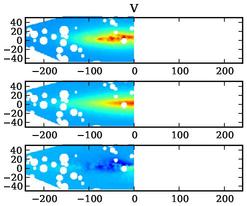}}
\caption[IC\,5052 - stellar fit (left) (1/2)]{Stellar decompositions for the left side of IC\,5052 (1/2). Each band consists of three panels. Top panels show the observation. Middle panels show the best-fit models. The lower panels show the difference maps. Colour scaling is equal between the three panels. The scale of the images is in arcsec.}\label{fig:stellar-IC5052_left1}
\end{figure*}

\begin{figure*}
\centering
\resizebox{0.48\textwidth}{!}{\includegraphics{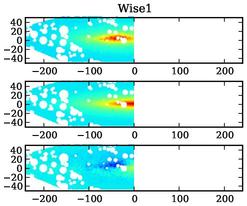}}
\resizebox{0.48\textwidth}{!}{\includegraphics{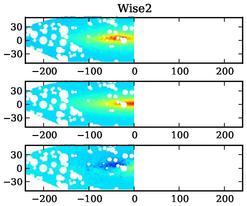}}
\caption[IC\,5052 - stellar fit (left) (2/2)]{Stellar decompositions for the left side of IC\,5052 (2/2). Each band consists of three panels. Top panels show the observation. Middle panels show the best-fit models. The lower panels show the difference maps. Colour scaling is equal between the three panels. The scale of the images is in arcsec.}\label{fig:stellar-IC5052_left2}
\end{figure*}

\begin{figure*}
\centering
\resizebox{0.48\textwidth}{!}{\includegraphics{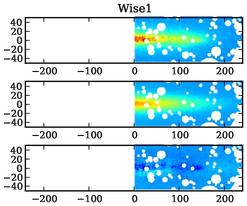}}
\resizebox{0.48\textwidth}{!}{\includegraphics{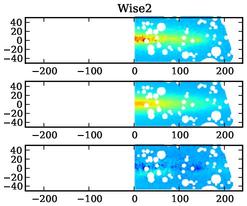}}
\caption[IC\,5052 - stellar fit (right) (1/2)]{Stellar decompositions for the right side of IC\,5052 (1/2). Each band consists of three panels. Top panels show the observation. Middle panels show the best-fit models. The lower panels show the difference maps. Colour scaling is equal between the three panels. The scale of the images is in arcsec.}\label{fig:stellar-IC5052_right1}
\end{figure*}

\begin{figure*}
\centering
\resizebox{0.48\textwidth}{!}{\includegraphics{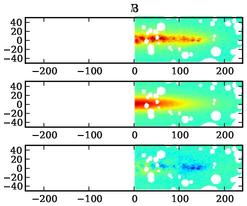}}
\resizebox{0.48\textwidth}{!}{\includegraphics{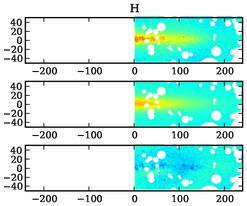}}
\resizebox{0.48\textwidth}{!}{\includegraphics{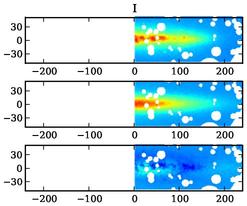}}
\resizebox{0.48\textwidth}{!}{\includegraphics{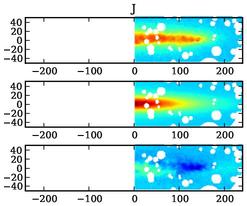}}
\resizebox{0.48\textwidth}{!}{\includegraphics{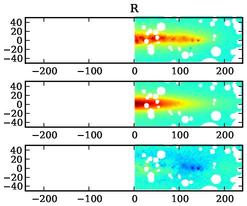}}
\resizebox{0.48\textwidth}{!}{\includegraphics{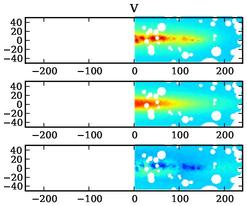}}
\caption[IC\,5052 - stellar fit (right) (2/2)]{Stellar decompositions for the right side of IC\,5052 (2/2). Each band consists of three panels. Top panels always show the observation. Middle panels show the best-fit models. The lower panels show the difference maps. Colour scaling is equal between the three panels. The scale of the images is in arcsec.}\label{fig:stellar-IC5052_right2}
\end{figure*}

\begin{figure*}
\centering
\resizebox{0.48\textwidth}{!}{\includegraphics{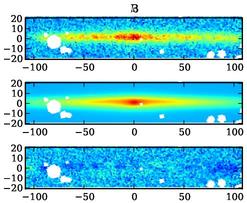}}
\resizebox{0.48\textwidth}{!}{\includegraphics{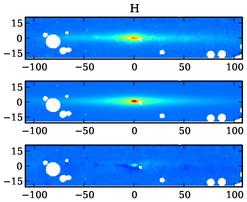}}
\resizebox{0.48\textwidth}{!}{\includegraphics{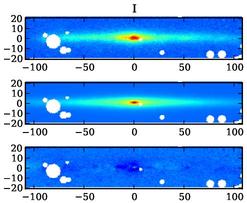}}
\resizebox{0.48\textwidth}{!}{\includegraphics{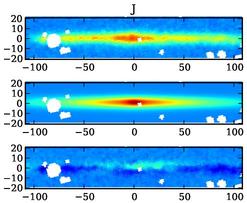}}
\resizebox{0.48\textwidth}{!}{\includegraphics{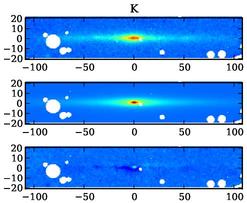}}
\resizebox{0.48\textwidth}{!}{\includegraphics{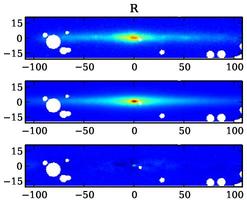}}
\caption[IC\,5249 - stellar fit (1/2)]{Stellar decompositions for IC\,5249 (1/2). Each band consists of three panels. Top panels always show the observation. Middle panels show the best-fit models. The lower panels show the difference maps. Colour scaling is equal between the three panels. The scale of the images is in arcsec.}\label{fig:stellar-IC5249a}
\end{figure*}

\begin{figure*}
\centering
\resizebox{0.48\textwidth}{!}{\includegraphics{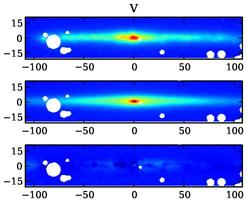}}
\resizebox{0.48\textwidth}{!}{\includegraphics{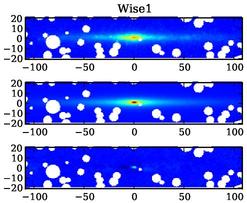}}
\resizebox{0.48\textwidth}{!}{\includegraphics{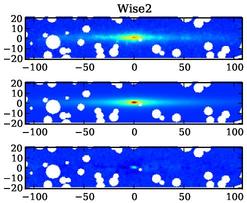}}
\caption[IC\,5249 - stellar fit]{Stellar decompositions for IC\,5249 (2/2). Each band consists of three panels. Top panels always show the observation. Middle panels show the best-fit models. The lower panels show the difference maps. Colour scaling is equal between the three panels. The scale of the images is in arcsec.}\label{fig:stellar-IC5249b}
\end{figure*}

\begin{figure*}
\centering
\resizebox{0.48\textwidth}{!}{\includegraphics{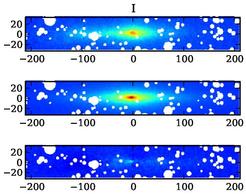}}
\resizebox{0.48\textwidth}{!}{\includegraphics{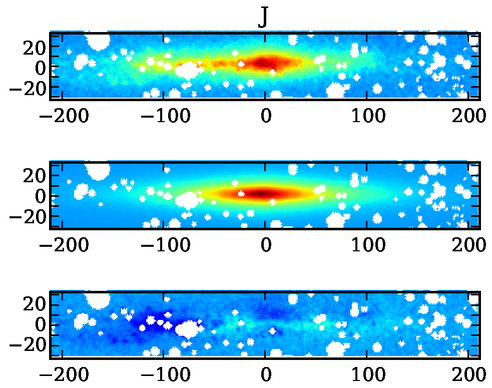}}
\resizebox{0.48\textwidth}{!}{\includegraphics{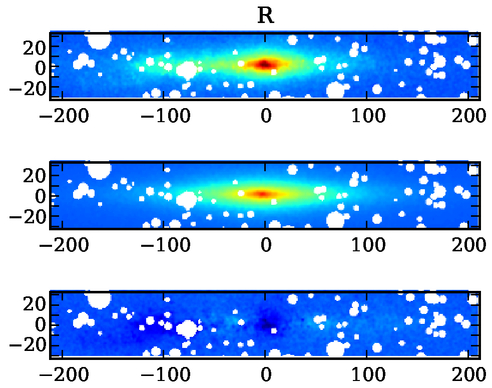}}
\resizebox{0.48\textwidth}{!}{\includegraphics{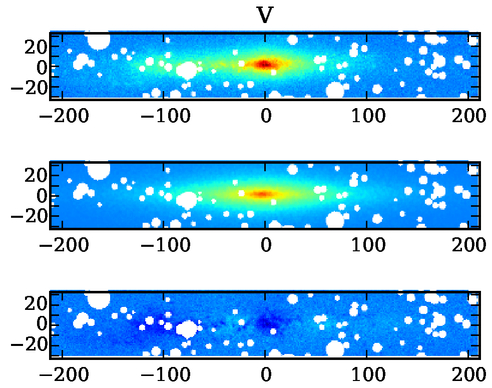}}
\resizebox{0.48\textwidth}{!}{\includegraphics{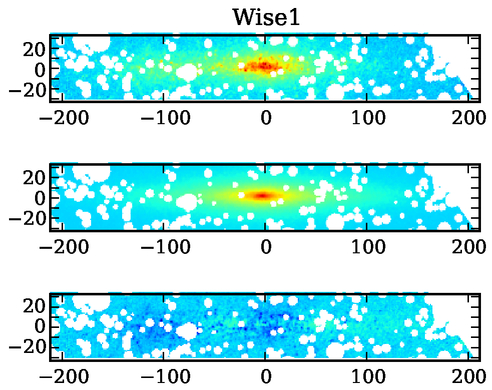}}
\resizebox{0.48\textwidth}{!}{\includegraphics{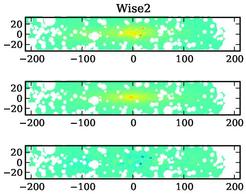}}
\caption[ESO\,115-G021 - stellar fit]{Stellar decompositions for ESO\,115-G021. Each band consists of three panels. Top panels always show the observation. Middle panels show the best-fit models. The lower panels show the difference maps. Colour scaling is equal between the three panels. The scale of the images is in arcsec.}\label{fig:stellar-ESO115-G021}
\end{figure*}

\begin{figure*}
\centering
\resizebox{0.48\textwidth}{!}{\includegraphics{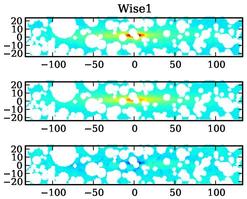}}
\resizebox{0.48\textwidth}{!}{\includegraphics{Petersetal-IVfig9f.jpg}}
\resizebox{0.48\textwidth}{!}{\includegraphics{Petersetal-IVfig9f.jpg}}
\resizebox{0.48\textwidth}{!}{\includegraphics{Petersetal-IVfig9f.jpg}}
\resizebox{0.48\textwidth}{!}{\includegraphics{Petersetal-IVfig9f.jpg}}
\resizebox{0.48\textwidth}{!}{\includegraphics{Petersetal-IVfig9f.jpg}}
\caption[ESO\,138-G014 - stellar fit]{Stellar decompositions for ESO\,138-G014. Each band consists of three panels. Top panels always show the observation. Middle panels show the best-fit models. The lower panels show the difference maps. Colour scaling is equal between the three panels. The scale of the images is in arcsec.}\label{fig:stellar-ESO138-G014}
\end{figure*}

\begin{figure*}
\centering
\resizebox{0.48\textwidth}{!}{\includegraphics{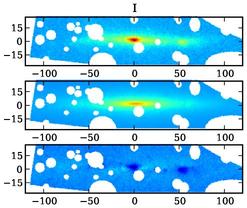}}

\resizebox{0.48\textwidth}{!}{\includegraphics{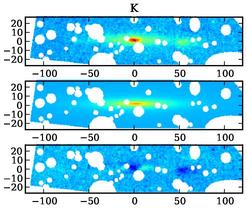}}
\resizebox{0.48\textwidth}{!}{\includegraphics{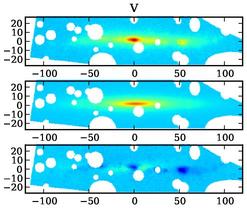}}
\caption[ESO\,146-G014 - stellar fit]{Stellar decompositions for ESO\,146-G014. Each band consists of three panels. Top panels always show the observation. Middle panels show the best-fit models. The lower panels show the difference maps. Colour scaling is equal between the three panels. The scale of the images is in arcsec.}\label{fig:stellar-ESO146-G014}
\end{figure*}

\begin{figure*}
\centering
\resizebox{0.48\textwidth}{!}{\includegraphics{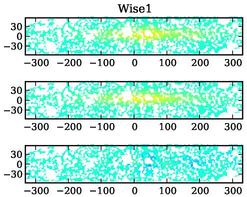}}
\resizebox{0.48\textwidth}{!}{\includegraphics{Petersetal-IVfig11d.jpg}}
\resizebox{0.48\textwidth}{!}{\includegraphics{Petersetal-IVfig11d.jpg}}
\resizebox{0.48\textwidth}{!}{\includegraphics{Petersetal-IVfig11d.jpg}}
\caption[ESO\,274-G001 - stellar fit]{Stellar decompositions for ESO\,274-G001. Each band consists of three panels. Top panels always show the observation. Middle panels show the best-fit models. The lower panels show the difference maps. Colour scaling is equal between the three panels. The scale of the images is in arcsec.}\label{fig:stellar-ESO274-G001}
\end{figure*}

\begin{figure*}
\centering
\resizebox{0.48\textwidth}{!}{\includegraphics{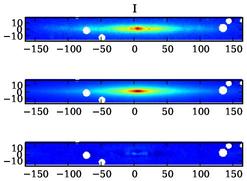}}
\resizebox{0.48\textwidth}{!}{\includegraphics{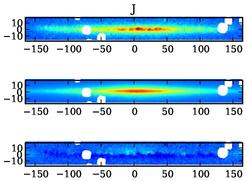}}
\resizebox{0.48\textwidth}{!}{\includegraphics{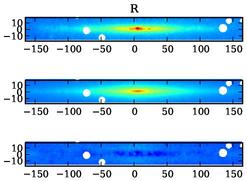}}
\resizebox{0.48\textwidth}{!}{\includegraphics{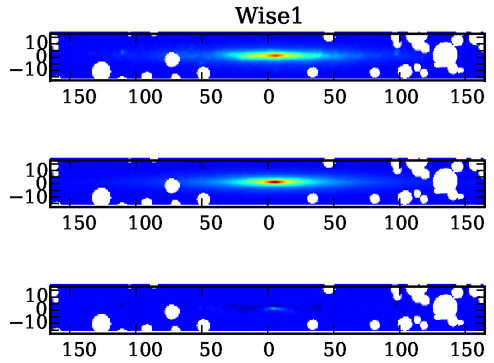}}
\resizebox{0.48\textwidth}{!}{\includegraphics{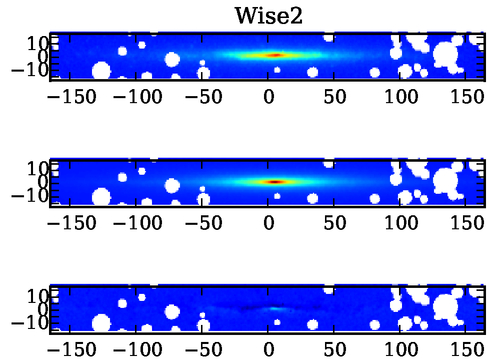}}
\caption[UGC\,7321 - stellar fit]{Stellar decompositions for UGC\,7321. Each band consists of three panels. Top panels always show the observation. Middle panels show the best-fit models. The lower panels show the difference maps. Colour scaling is equal between the three panels. The scale of the images is in arcsec.}\label{fig:stellar-UGC7321}
\end{figure*}

\section{Data Fitting}\label{section:stellardatafitting}
The data will be modelled using \textsc{FitSKIRT} with the following 
components \citep{Geyter2013A}.
The disc of each galaxy is modelled as a double-exponential disc.
No truncation was included in the model, as this is currently not 
supported by \textsc{FitSKIRT}.
Truncations are known to occur near the onset of the \hi warp 
\citep{vanderKruit2007A} and this region is not of interested 
for our goal of measuring the mid-plane hydrostatics (see also 
Section 2 of Paper III).
Developing a truncated model was discussed, but considering the 
quality of the data, the current research questions and the 
computational cost associated with additional free parameters, 
we have decided not to develop such an option.
The disc thus follows the luminosity distribution
\begin{equation}
 j(R,z) = \frac{L_{\textrm{d},*}}{4 \pi h^2_{R,*} h_{z,*}} \exp{\left(-\frac{R}{h_{R,*}}\right)}\exp{\left(-\frac{|z|}{h_{z,*}}\right)}\,\,,\label{eqn:stellardistribution}
\end{equation}
with $L_{\textrm{d},*}$ the total disc luminosity, $h_{R,*}$ and $h_{z,*}$ the disc 
scale length and scale height, and $R$ and $z$ as the radial and vertical 
coordinates.

The assumption of a double-exponential model in our fits is a 
simplification of reality. It is well-known that better fits are possible;
\citet{vdkruit88} proposed a sech-function as a compromise between the 
isothermal sech${^2}$ distribution of \citet{vdks1981}, which ignores the 
real situation that there are most likely age groups of stars with different 
velocity dispersions, and the mathematically convenient, but unphysical 
exponential. We believe our fits are sufficiently satisfactory that any further
sophistication is not necessary, also in view of the fact that at low $z$, 
where the exponential and the sech deviate most, dust extinction would need to 
be fitted more realistically then the crude way we apply here, 
namely also with a double exponential. Also the 
scaleheight of the stars may very well change with galactocentric distance, 
as found by \citet{RichReyn} and \citet{NJ2002}. Again our fits work very well
with a constant scaleheight, and we therefore believe these are adequate for 
our purposes. We return to this issue in section 5.2.

Our galaxies are close to edge-on and already \citet{vdks1981} showed
that small variation in invclination away from perfectly edge-on do not affect
the determined scale height. Nevertheless, we leave the inclination in as a
parameter (see below).

A bulge was also included, which follows the luminosity density 
\citep{Geyter2013A}
\begin{eqnarray}
j(R,z) &=& \frac{L_{\textrm{b},*}}{q R_\textrm{e}^3} S_\textrm{n} \left(\frac{m}{R_\textrm{e}}\right)\,\,,\\
m &=& \sqrt{R^2+\frac{z^2}{q^2}}\,\,,\label{eqn:bulgedistribution}
\end{eqnarray}
with the total bulge luminosity $L_{\textrm{b},*}$, the effective 
radius $R_e$, the S\'ersic function 
$$
S_\textrm{n} = I_e\exp -C \left[ \left( {R \over R_e } \right)^{1/n} - 1 \right]
$$ 
(with the proper normalization) and the flattening 
of the bulge $q$. In general our photometry is not deep enough to look
for thick disks and what we fit as bulges may very well be the brighter 
parts of suchgcomponents.

The dust is also modelled as a double-exponential disc
\begin{equation}
 \rho_\textrm{d}(R,z) = \frac{M_\textrm{d}}{2 \pi h^2_{R,d} h_{z,d}} \exp{\left(-\frac{R}{h_{R,d}}\right)}\exp{\left(-\frac{|z|}{h_{z,d}}\right)}\,\,,
\end{equation}
with the dust density $\rho$, the total dust mass $M_\textrm{d}$, 
and the dust scale length and scale height $h_\textrm{R,d}$ and $h_\textrm{z,d}$.
The galaxies in our sample are late-type dwarfs and we do not 
expect them to have a lot of dust.
We include dust in the results for completeness, but stress that 
we are skeptical about the exact quantities returned by the fitting.
The inclination can vary between $86^\circ$ and $90^\circ$ to get 
the best fit. However, with the absence of dust bands in most 
of our galaxies, we cannot confirm the exact inclination.

Including the inclination $i$ and the exact central position, 
the model has a total of 11 global free parameters, plus two 
additional free parameters per band (the disc luminosity 
$L_{\textrm{d},*}$ and bulge-to-disc ratio $B/D$).
\textsc{FitSKIRT} fits the data using the \textsc{GaLib} 
genetic algorithm \citep{galib}.
We use a population of 200 individuals and evolve the model 
for 100 iterations.
We run five fits for each galaxy.
The Millipede Cluster of the University of Groningen was used 
to perform all fits. A fit typically takes 30 hours.

Due to the random path that the light follows in the model, no two 
instances of the model will be the same. As such, both the 
observation and the model contain Poisson distributed noise.
The objective function to measure the error is therefore 
\begin{eqnarray}
\chi^2  &=& \sum_j^{N_\textrm{pix}}{\frac{\left(I_{\textrm{mod},j}  - I_{\textrm{obj},j}\right)^2   }{\sigma^2}}\,\,,\\
       &=& \sum_j^{N_\textrm{pix}}{\frac{\left(I_{\textrm{mod},j}  - I_{\textrm{obj},j}\right)^2 }{|I_{\textrm{obj},j}| + I_{\textrm{mod},j}}}\,\,.
\end{eqnarray}
The lowest $\chi^2$ value is adopted as the true value. The error in each 
parameter is taken as the standard deviation from the five runs.

\section{Results}\label{sec:stellarresults}

The results for each galaxy are shown in Table \ref{tbl:stellar-main1}.
The quality of the individual fits can be seen in Table 
\ref{tbl:stellar-main2} in the online Appendix. 
When available, we also list the published integrated 
magnitude $L_{\textrm{band,*}}$. 
Where needed, these have been corrected for Galactic 
foreground extinction using the \citet{reddeningA} 
re-calibration of the \citet{reddeningB} infrared-based 
dust map, as calculated in NED. 
Before the disc luminosity $L_{\textrm{d},*}$ produced in \textsc{FitSKIRT} can 
be used for further analysis, it first needs to be calibrated.
This can be done by multiplying it with $10^{15} L_{\textrm{band,*}}$, where 
$L_{\textrm {band,*}}$ is the integrated luminosity, expressed in W/m$^2$, as 
measured without dust absorption.

\subsection{IC\,2531}
Galaxy IC\,2531 is the only Sb galaxy in our sample, and has the highest 
maximum circular velocity at $v_\textrm{max}=260.5$\,km/s 
(Table \ref{tbl:stellar-main1}).
This is nearly twice the maximum circular velocity of the other galaxies 
in our sample.
The central brightness component has not been reproduced correctly. 
This bulge is most likely peanut-shaped or boxy, which requires a 
different bulge model than adopted here \citep{Jarvis1986A,deSouza1987A}.
The scale length is $h_{\textrm{R},*}=5077\pm262$\,pc and the scale height 
 $h_{\textrm{z},*}=613\pm46$\,pc.
\citet{Kregel2002} report a much longer $h_{\textrm{R},*}=12511\pm2643$\,pc 
and $h_{\textrm{z},*}=658.5\pm65.8$\,pc.
The galaxy features a very prominent dust lane, which Kregels I-band model 
does not treat for and so their scale length might be overestimated.
\citet{Xilouris1999} has also performed a fit to the galaxy that did 
include a treatment for the dust. 
They report a K-band scale length of  $h_{\textrm{R},*}=5.04\pm0.1$\,kpc 
and a scale height of $0.45\pm0.02$\,kpc.
The dust is found to have a (K-band) scale length of 
$h_{\textrm{R},d}=8.00\pm0.3$\,kpc and a scale height of
 $h_{\textrm{z},d}=0.22\pm0.03$\,kpc, which is similar to the 
$h_{\textrm{R},d}=7.0\pm2.2$\,kpc and  $h_{\textrm{z},d}=0.3\pm0.1$\,kpc reported here.

\subsection{IC\,5052}
Galaxy IC\,5052 has proven hard to model, as the galaxy showed clear 
asymmetries.
We therefore opted to model both sides of the galaxy separately 
(Figures \ref{fig:stellar-IC5052_left1}, \ref{fig:stellar-IC5052_left2}, 
\ref{fig:stellar-IC5052_right1} and \ref{fig:stellar-IC5052_right2}).
This forced us to fix the centre of the galaxy, which has clearly hampered 
the results.
From an inspection of the images, we conclude that the left side better 
represents the galaxy, although the fit is far from perfect. 
The scale length for the left side $h_{\textrm{R},*}$ is $857\pm68$\,pc and 
the scale height  $h_{\textrm{z},*}$ is $124\pm80$\,pc, which makes this the 
galaxy with the shortest scale length in our sample.
\citet{Comeron2011B} has also modelled this galaxy using a double disc 
approach, and report a thick disc scale length of 470-530\,pc and a thin 
disc scale length of 140-170\,pc. 
They note that the galaxy can also be modelled successfully with a single disc.
Overall, we find that the galaxy is very clumpy in its light distribution, 
which cannot be fit properly by the models. Our fitted scalelength
differs considerably between the two sides, and both are much larger than the
determined value of \citet{Comeron2011B}. Our result should be treated 
with much cuation. This galaxy proven to be too 
difficult to model for our purposes and it has been deleted from our
sample in Paper V, where we do the final analysis.

\subsection{IC\,5249}
The overall quality of the fit is very good.
The scale length is $h_{\textrm{R},*}=6828\pm340$\,pc, while the scale height 
is $h_{\textrm{z},*}=242\pm20$\,pc.
Similar results have also been found in previous studies.
\citet{Carignan1983} analysed the galaxy in the B-band and distinguished 
two components wityh scale lengths 18\,kpc and 2.5\,kpc.
\citet{Wainscoat1986} confirmed these results using H, I and K-band data.
The galaxy was also analysed by \citet{Byun1992} in the B, R and I 
\citep{vanderKruit2001A}.
\citet{Abe1999} discovered a very sharp truncation at two scale lengths.
We do not find any indication of a second disc component as 
\citet{Carignan1983}, although their second component might 
well have been the bulge we find here. This galaxy is an example of a so-called superthin galaxy. \citet{vanderKruit2001A}, who found a scale lenght of about 7 kpc attributed this to the relatively long scale length combined with the low (face-on) surface brightness.

\subsection{ESO\,115-G021}
A large amount of masking was required for ESO\,115-G021 (Figure 
\ref{fig:stellar-ESO115-G021}).
The dwarf galaxy has been modelled successfully, although it is still clumpy.
The stellar disc scale length $h_{\textrm{R},*}$ is only $1108\pm280$ pc, 
while the scale height $h_{\textrm{z},*}$ is only $149$ parsec. 
The galaxy is very faint and emits only $5.0\times10^7$L$_\odot$ in the R-band. 
In comparison to the neutral hydrogen mass, the $M_\textrm{HI}/L_\textrm{R}$ 
is 12.5, the highest in our sample.

\subsection{ESO\,138-G014}
ESO\,138-G014 also required extensive masking (Figure 
\ref{fig:stellar-ESO138-G014}).
Despite this, the overall fit is good. 
With  $h_{\textrm{R},*}=2288\pm59$\,pc and $h_{\textrm{z},*}=217\pm10$\,pc, 
the stellar disc is twice as long as ESO\,115-G021 and slightly thicker. 
\citet{Kregel2002} previously modelled this galaxy without dust and 
reported a larger disc with $h_{\textrm{R},*}=3779$\,pc and 
$h_{\textrm{z},*}=382\pm10$\,pc.

\subsection{ESO\,146-G014}
ESO\,146-G014 is a known low-surface brightness galaxy and 
is extremely metal poor \citep{Roennback1995,Morales2011A}. 
The galaxy is a slow rotator $v_\textrm{max}=84.1$\,km/s and 
is very patchy in nature, as can be seen in Figure 
\ref{fig:stellar-ESO146-G014}.
Because of this, the fit is far from perfect. 
As can be seen in all three bands, the galaxy has a 
very bright central region, but also has a bright 
spot 50'' east of the centre. 
The galaxy is also asymmetric, as can be seen most clearly in the V-band image.

\subsection{ESO\,274-G001}
Galaxy ESO\,274-G001 is very close to the Galactic plane, with a Galactic 
latitude of only $9.3^\circ$.
The four bands therefore required an exceptional amount of masking (Figure 
\ref{fig:stellar-ESO274-G001}), but the fit was still successful.
The galaxy has a stellar disc scale length $h_{\textrm{R},*}$ of $1270\pm59$\,pc 
and a scale height $h_{\textrm{z},*}$ of $161\pm2$\,pc, which is very similar to 
ESO\,115-G021.
The overall appearance is different from that galaxy, as ESO\,274-G001 has a 
far more flattened bulge.
Similar to the ESO\,115-G021 and ESO\,146-G014, the galaxy is a slow rotator, 
with a maximum circular velocity of $v_\textrm{max}=103.9$\,km/s.
Moreover, similar to those two galaxies, ESO\,274-G001 is  patchy in nature.
The galaxy has the highest luminosity compared to its optically thin HI mass: 
$M_\textrm{HI} / L_\textrm{R}=0.8$

\subsection{UGC\,7321}
UGC\,7321 has been modelled using five bands (Figure \ref{fig:stellar-UGC7321}),
 the results of which have the lowest combined $\chi^2$ error in this sample.
The scale length has been measured at $h_{\textrm{R},*}=2498\pm349$\,pc and the 
scale height at  $h_{\textrm{z},*}=187\pm80$\,pc. 
This is in agreement with \citet{OBrien2010D}, who applied the deprojection 
method and reported $h_{\textrm{d},*}=2650$\,pc and $h_{\textrm{z},*}=245$\,pc. 
The scale-height was reported as 140-150\,pc by \citet{Matthews1999A}, who 
did not discern a bulge. We emphasize that our fitted bulge may also 
correspond to the brighter parts of a thick disk,

\citet{OBrien2010D} performed a rotation-curve decomposition on the galaxy 
and found a mass to light $M_*/L_{r'}$ upper limit of 1.05.
This implies a recent burst of star formation in the galaxy, which matches 
the detection of a significant fraction of young stars in the disc 
\citep{Matthews1999A}.
\cite{OBrien2010D} also performed a fit to the vertical hydrostatics 
of the disc and found a good fit at $M_*/L_R$ of 0.2.

\section{Discussion}\label{sec:stellardiscussion}
\subsection{Quality of the Fits}
What can we conclude about the overall quality of the fits?
Comparing to the available literature, we see that IC\,2531, IC\,5249 
and UGC\,7321 are in agreement with the results by others. 
The slower rotating galaxies, e.g. ESO\,115-G021, ESO\,146-G014 and 
ESO\,274-G001 all have a clumpy nature, which makes fitting harder.
For IC\,5052 and ESO\,138-G014, the parameters reported by other authors 
are different. As we noted before, IC\,5052 gave us problems to model, 
so our results could well be wrong. 
ESO\,138-G014 also required extensive masking.
It is a low surface brightness galaxy, but the  fit we have performed 
still looks acceptable (see Figure \ref{fig:stellar-ESO138-G014}).
As such, we feel that the derived stellar disc scale lengths and scale 
heights are reasonably accurate, with the possible exception of IC\,5052.

A similar endeavor as the work done in this paper has been previously 
undertaken by \citet{deGeyter2014}, who set out to model edge-on galaxies 
in the Calar Alto Legacy Integral Field Area
Survey (CALIFA), which showed a clear dust band.
In total they settled on a sample of 12 --mostly early-type-- galaxies for 
which they performed an oligochromatic fit to the $g'$, $r'$, $i'$ and $z'$ bands.
While their data is of higher quality, their results are very similar to 
ours, and they ``conclude that in general most galaxies
are modelled accurately, especially when keeping in mind that the
FitSKIRT models only consist of three basic components and they
were determined by an automated procedure over a large parameters
space without strong initial boundary conditions'' \citep{deGeyter2014}.

\begin{table*}
\resizebox{0.98\textwidth}{!}{
\centering
    \begin{tabular}{l|cc|ccc|cc|cc|c}
  ~         &{Stellar } &{Stellar } &{Dust } &{Dust } &{Dust} &{Bulge } &{Bulge } &~& S\'ersic \\  
  {Name}         &{ scale length} &{scale height} &{scale length } &{ scale height} &{mass} &{ Radius } &{ flattening} &{ Inclination}~&{Index} & QOF\\
    ~         &[pc] &[pc] &[pc] & [pc] &[$10^7$ M$_\odot$] &{ [pc]} & & $^\circ$ & ~  \\\hline\hline
IC\,2531 & $5077\pm262$ & $613\pm46$ & $6997\pm2213$ & $311\pm109$ & $4\pm1.3$ & $5451\pm1775$ & $0.78\pm0.22$ & $88.7\pm0.8$ & $5.6\pm1.0$ & G\\
IC\,5052 (left) & $857\pm68$ & $124\pm80$ & $1500\pm3655$ & $600\pm15$ & $0.5\pm0.44$ & $1600\pm518$ & $0.25\pm0.02$ & $86.0\pm1.7$ & $0.5\pm0.2$ & M \\
IC\,5052 (right) & $1617\pm129$ & $781\pm299$ & $6664\pm2042$ & $523\pm96$ & $0.01\pm0.5$ & $2149\pm45$ & $0.16\pm0.05$ & $87.7\pm1.0$ & $0.5\pm0.0$ & B \\
IC\,5249 & $6828\pm340$ & $242\pm20$ & $4380\pm677$ & $372\pm67$ & $0.88\pm0.2$ & $4602\pm183$ & $0.22\pm0.01$ & $88.8\pm0.9$ & $2.9\pm0.2$ & G \\
ESO\,115-G021 & $1108\pm280$ & $149\pm101$ & $10515\pm2057$ & $432\pm69$ & $2.4\pm0.47$ & $3810\pm1159$ & $0.35\pm0.06$ & $86.0\pm0.7$ & $2.6\pm0.7$ & G\\
ESO\,138-G014 & $2288\pm59$ & $217\pm10$ & $3074\pm508$ & $509\pm34$ & $1.5\pm0.23$ & $5843\pm611$ & $0.18\pm0.01$ & $86.8\pm0.3$ & $3.7\pm0.3$ & G\\
ESO\,146-G014 & $5356\pm1384$ & $1000\pm422$ & $7115\pm1063$ & $525\pm58$ & $2.3\pm0.47$ & $3264\pm1356$ & $0.09\pm0.10$ & $86.0\pm0.0$ & $0.9\pm0.3$ & G \\
ESO\,274-G001 & $1270\pm59$ & $161\pm2$ & $11374\pm373$ & $472\pm33$ & $2.9\pm0.69$ & $4880\pm551$ & $0.06\pm0.02$ & $86.8\pm0.2$ & $0.6\pm0.1$ & G\\
UGC\,7321 & $2498\pm349$ & $187\pm80$ & $8568\pm1956$ & $600\pm139$ & $6.2\pm1.1$ & $2532\pm1013$ & $0.10\pm0.02$ & $90.0\pm1.1$ & $1.9\pm0.7$ & G \\\hline\hline
    \end{tabular}
}
    \caption[Global properties of the stellar fits]{The global properties of the stellar fits per galaxy. QOF is the quality of the fit: Good, Moderate or Bad.}\label{tbl:stellar-main1}
\end{table*}

% Onderdeel van de discussie
\subsection{On the $z$-distribution}
A key assumption in our model is the use of a double-exponential disk (see Equation 1).
This choice was made as a basis for the hydrostatic equilibrium calculations in Paper V.
Other functional forms for the vertical distribution were available, but were rejected as the quality of the observations does not allow us to distinguish accurately enough between the various forms.

To demonstrate this, we have present the vertical distribution of UGC\,7321 along a slice of the galaxy in Figure 13. 
UGC\,7321 was one of our most succesful fits, and as such the profiles between the model and the observation match sufficiently well.
But even in this slice, there is difference in the quality of the fit. 
The Wise2 fit has been very well reproduced.
Yet the R-band profile has been reproduced considerably less well, showing excess light futher above and beyond the plane.
The fits are influenced by the fact that the program tries to compensate for the local structures and contaminations (e.g. stray light due to field stars).
More accurate and less contaminated observations are required to accurately test which functional form of the vertical distribution works best.

\begin{figure*} % Figure 1
\centering
\resizebox{0.32\textwidth}{!}{\includegraphics{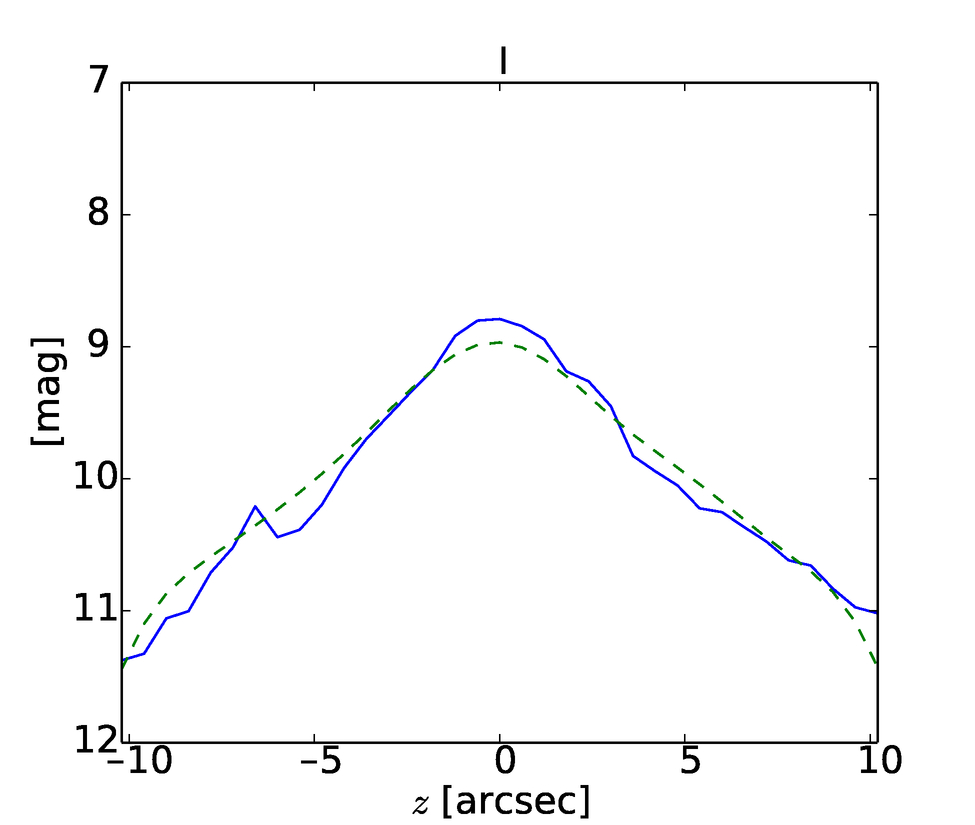}}
\resizebox{0.32\textwidth}{!}{\includegraphics{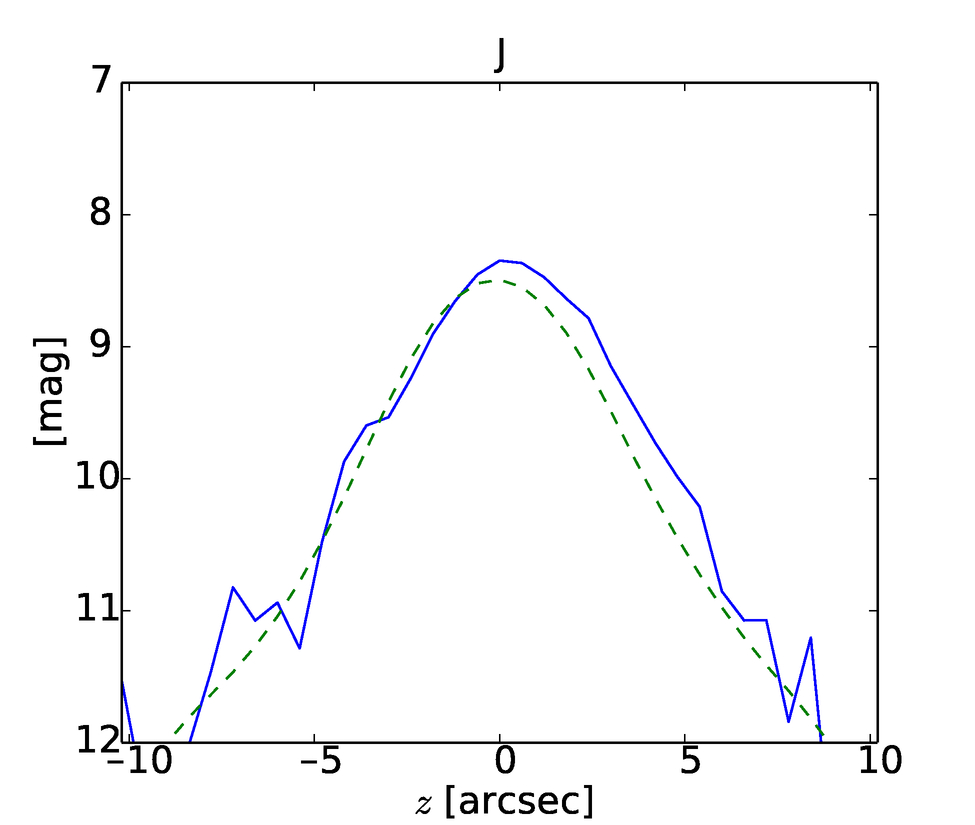}}
\resizebox{0.32\textwidth}{!}{\includegraphics{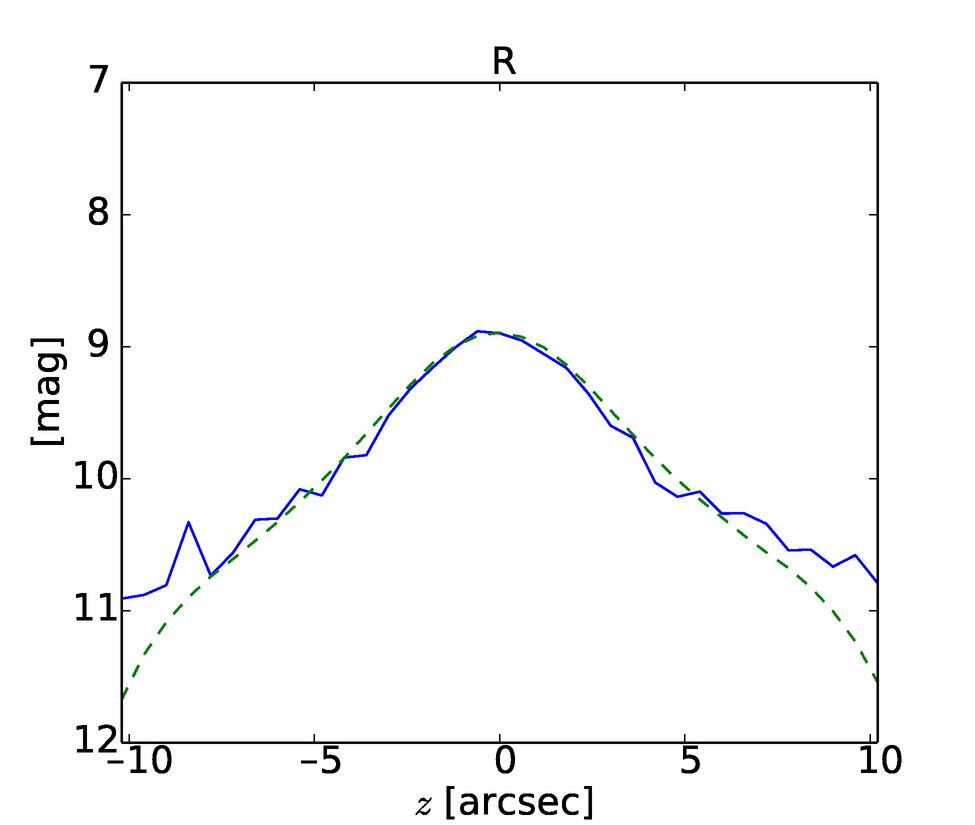}}
\resizebox{0.32\textwidth}{!}{\includegraphics{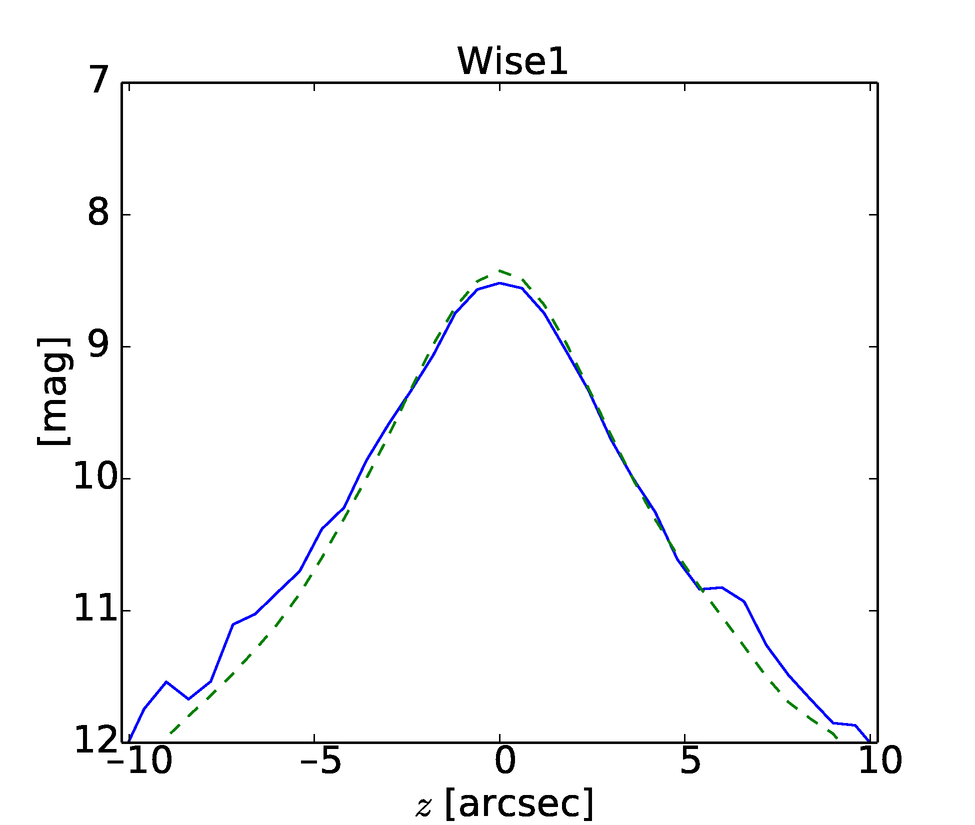}}
\resizebox{0.32\textwidth}{!}{\includegraphics{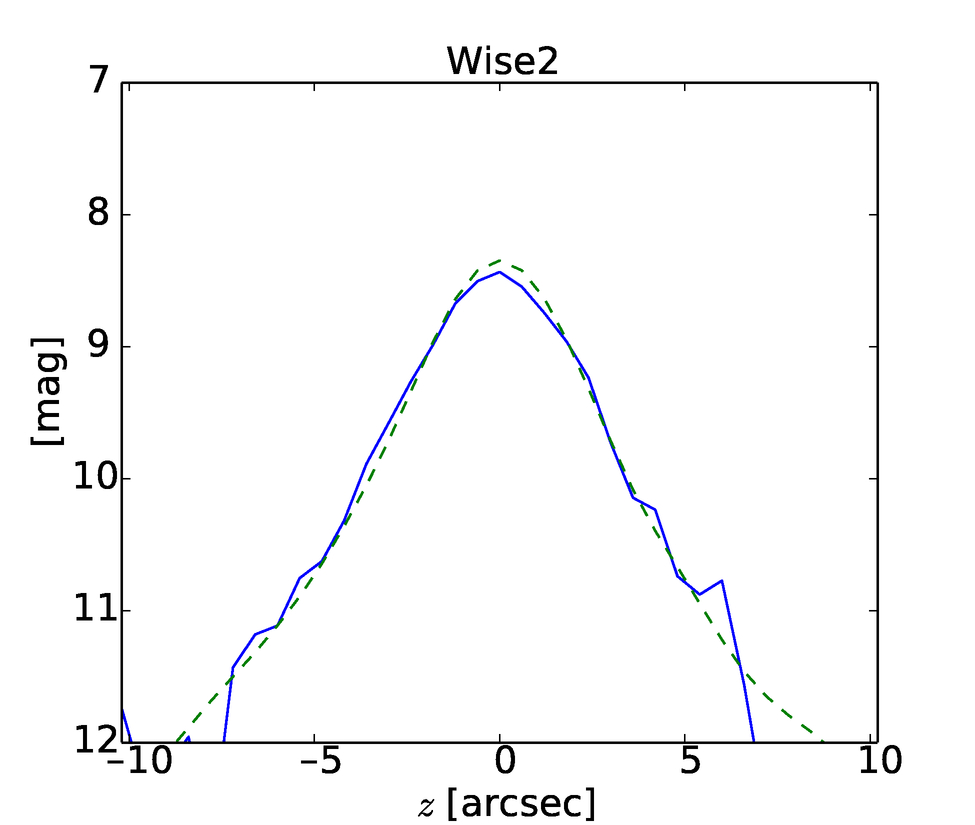}}
\caption{Profile of a vertical slice of UGC\,7321. The profile was taken at a distance of 28.4$^{\prime\prime}$ from the center of the galaxy. Solid lines represent the observation. Dashed lines the simulated galaxy. Zeropoint is uncalibrated.}
\end{figure*}

\subsection{Discs and Bulges}
Let us now compare  the global properties of the stellar discs.
Focusing on the seven good fits, the average stellar disc scale length in 
our sample is $h_{R,*}=3.49\pm2.24$\,kpc, while the average scale height of 
the stellar disc is $h_{z,*}=0.37\pm0.32$\,kpc. 
As a comparison \citet{deGeyter2014} report $4.23\pm1.23$\,kpc and 
$0.51\pm0.27$\,kpc for their 12 edge-on sample.
They also calculate the mean values for the 34 edge-on galaxies in 
\citet{kkg02}, reporting $4.73\pm2.57$\,kpc and scale height of 
$0.57\pm0.25$\,kpc.
That sample consists of Sa, Sb and Sc type galaxies, while ours 
consists of mostly of Sd type galaxies (Table 2 of Paper I of this series).
IC\,5249 appears to be a unique galaxy in terms of scale length 
to scale height ratio. 
If we remove it from the sample, the averages becomes even more distinct with 
$h_{R,*}=2.93\pm1.85$\,kpc and $h_{z,*}=0.38\pm0.28$\,kpc.
It is typically not expected that the scale lengths for disc galaxies depend
on Hubble morphological types between Sa and Sc \citep{dj96b, Graham2001A}.
However, for Scd and Sd galaxies, it was demonstrated using face-on galaxies 
that the scale length tends to be significantly shorter \citep{Fathi2010} and 
our results are within expectations, although the range we find is large,
the mean is still less than the values we just quoted for Sa to Sc galaxies.

If we now focus on the scale length over scale height ratio, we find an 
average of $h_{R,*}/h_{z,*}=11.6\pm7.76$. 
IC\,5249 has the by far highest value at 28.2, which makes it a `super 
thin' disc. 
Removing this one galaxy, the average ratio becomes $h_{R,*}/h_{z,*}=8.81\pm2.78$.
A very similar value of $h_{R,*}/h_{z,*}=8.26\pm3.44$ was reported by 
\citet{deGeyter2014} for their own sample, and a value of 
$h_{R,*}/h_{z,*}=8.21\pm2.36$ for the \citet{kkg02} sample.
We conclude that our measurements are, with the 
possible exception of IC\,5249, in good agreement with their samples.

Looking at the bulges, we find a typical bulge effective radius of 
$R_e=4.34\pm1.19$\,kpc.
This is longer than reported by \citet{deGeyter2014}, who reports 
the value of $R_e=2.31\pm1.59$\,kpc.
This was already large compared to the 1000 galaxies sample of 
\citet{Gadotti2009}, who reported $R_e=0.84\pm0.36$\,kpc.
It was argued by \citet{deGeyter2014} that this difference is 
due to the lack of dust attenuation correction in \citet{Gadotti2009}, 
and due to the lack of a treatment of bars in their own work.
The \citet{Gadotti2009} sample does not contain many Sd galaxies; there 
might be different averages that apply for an Sd sample.
However, given our results from the previous section, we argue that it is 
more likely that our bulges are not fitted very reliably.
The average S\'ersic index we find is $2.6\pm1.7$, which is consistent with 
the $2.37\pm1.35$ reported by \citet{deGeyter2014}. As we
poinbted out above, there is a real 
possibility that our bulge components really are the brighter parts of a thick 
disk.

\subsection{Dust}
Most of the galaxies in our sample are slow rotators. 
The Sd type galaxies have a maximum circular rotation of 
$v_\textrm{max}=131.9$\,km/s. 
Only IC\,2531 has a much higher rotation of $v_\textrm{max}=260.5$\,km/s 
(Table 4 of Paper I). 
\citet{Dalcanton2004} found that for galaxies rotating slower than a 
circular velocity of $v_\textrm{max}\!\sim\!120$\,km/s, no dust lane forms. 
Instead, the dust settles in more clumpy structure.
Inspecting the images of the galaxies, we can confirm this distinction as 
well in our own sample, although it is hard to distinguish for the discs 
just above 120\,km/s. 
The average dust scale length and scale height are $h_{R,d}=4.34\pm1.19$\,kpc 
and $h_{z,d}=0.46\pm0.01$\,kpc, while the ratio between the two is on average 
$h_{R,d}/h_{z,d}=10.1\pm4.4$. 
A typical dust scale length and height of $h_{R,d}=6.03\pm2.92$ and 
$h_{z,d}=0.23\pm0.10$ was reported by \citet{deGeyter2014}.
For the scale height similar values of $h_{z,d}=0.23\pm0.08$\,kpc and 
$h_{z,d}=0.25\pm0.11$ were reported by \citet{Xilouris1999} and 
\citet{Bianchi2007}.
Our galaxies thus typically have a thicker dust layer than the 
samples of these authors.

\citet{Dalcanton2004} argued that for discs  rotating slower than 
a circular velocity of $v_\textrm{max}\!\sim\!120$\,km/s the dust 
would follow a thicker distribution than the faster rotating sample.
Indeed, most of our galaxies are below or near a circular velocity 
of 120\,km/s.
Taking the three galaxies with circular velocities above 130\,km/s, 
we have an average scale length to scale height ratio of 
$h_{R,d}/h_{z,d}=13.8\pm3.3$, compared to $h_{R,d}/h_{z,d}=7.4\pm2.7$ 
for the other four. 
The sample of \citet{deGeyter2014} has a mean ratio of $h_{R,d}/h_{z,d}=26.2$, 
which is much closer to our fast rotating galaxy at $h_{R,d}/h_{z,d}=17.5$.

However, UGC\,7321 with a maximum circular velocity of 
$v_\textrm{max}=128$\,km/s has a very different ratio of $h_{R,d}/h_{z,d}=4.2$. 
This galaxy also has the most dust mass of the sample, 
with $6.2\times10^7$\,M$_\odot$, compared to the average of 
$2.8\pm1.82\times10^7$\,M$_\odot$ for the entire sample. 
There is no distinction in terms of dust mass to be made 
between the slow and quick rotators.

The dust scale length to stellar scale length ratio is on 
average $h_{R,d}/h_{R,*}=1.88\pm1.37$ in our sample. 
This is compatible to the $h_{R,d}/h_{R,*}=1.73\pm0.83$ reported 
by \citet{deGeyter2014}.
\citet{Xilouris1999} reports $h_{R,d}/h_{R,*}=1.36\pm0.17$ and 
\citet{Bianchi2007} finds $h_{R,d}/h_{R,*}=1.53\pm0.55$, both values 
are compatible with ours.
The dust scale height to stellar scale height ratio we report is 
$h_{z,d}/h_{z,*}=1.99\pm1.15$.
Although compatible, this is much higher than the value of 
$h_{z,d}/h_{z,*}=0.55\pm0.22$ reported by \citet{deGeyter2014}, the 
$h_{z,d}/h_{z,*}=0.58\pm0.13$ reported by \citet{Xilouris1999} and the 
$h_{z,d}/h_{z,*}=0.52\pm0.49$ reported by \citet{Bianchi2007}.
We thus find that the dust to stellar scale length ratio of our 
late-type galaxies sample is compatible to more early-type samples; 
the dust to stellar scale height ratio is far higher. 
The dust in our slow rotating galaxies sample forms into a much thicker 
disc, as predicted by \citet{Dalcanton2004}.

\section{Conclusions}\label{sec:stellarconclusions}
In this paper, we have attempted to  fit the bulge and disc of eight edge-on 
dwarf galaxies using \textsc{FitSKIRT} automatically.
The quality of our fit varies, mostly due to limited quality of the available 
data and the intrinsic low luminosity of these galaxies.
Despite this, we have successfully recovered the stellar discs in seven out 
of eight of the galaxies in our sample.
The results for the bulges are less reliable, which is most likely 
due to a lack of an accurate description for bars in the code.
We have also successfully measured the dust distribution in these seven 
galaxies.
The average dust scale length to stellar scale length is compatible with 
other samples, but our dust scale height to stellar scale height ratio is 
far higher than in typical other samples.
The HI mass to light ratio $M_\hi/L_R$ varies drastically between the 
various galaxies. 
It is only 0.2 in ESO\,274-G001, yet 12.5 in ESO\,115-G021.

\section*{Acknowledgments}
SPCP is grateful to the Space Telescope Science Institute, Baltimore, USA, the 
Research School for Astronomy and Astrophysics, Australian National University, 
Canberra, Australia, and the Instituto de Astrofisica de Canarias, La Laguna, 
Tenerife, Spain, for hospitality and support during  short and extended
working visits in the course of his PhD thesis research. He thanks
Roelof de Jong and Ron Allen for help and support during an earlier 
period as visiting student at Johns Hopkins University and 
the Physics and Astronomy Department, Krieger School of Arts and Sciences 
for this appointment.

PCK thanks the directors of these same institutions and his local hosts
Ron Allen, Ken Freeman and Johan Knapen for hospitality and support
during many work visits over the years, of which most were 
directly or indirectly related to the research presented in this series op 
papers.

Work visits by SPCP and PCK have been supported by an annual grant 
from the Faculty of Mathematics and Natural Sciences of 
the University of Groningen to PCK accompanying of his distinguished Jacobus 
C. Kapteyn professorhip and by the Leids Kerkhoven-Bosscha Fonds. PCK's work
visits were also supported by an annual grant from the Area  of Exact 
Sciences of the Netherlands Organisation for Scientific Research (NWO) in 
compensation for his membership of its Board.

%\begin{thebibliography}{99}

%\setlength{\bibsep}{0.1em}
\bibliography{refsIV}
\bibliographystyle{mn2e}

%\end{thebibliography}

%\onecolumn

\appendix

\section{Tabular material}
The following pages show the details of the observations in Table 
\ref{tbl:stellarobservations} and the $\chi^2$ 
quality of the fit for each filter band in  Table 
\ref{tbl:stellar-main2}.

\begin{table*}
\centering
%\resizebox{0.98\textwidth}{!}{
\begin{tabular}{lclll}
Galaxy          & Filter    & Telescope                 & Observer/Project      & Date\\\hline\hline
IC\,2531         & I         & ANU 40inch Telescope      & J.C. O'Brien          & 02-03-2002\\
                & R         & ESO\,La Silla              &  ESO(B) Atlas         & 04-02-1986\\
                & $3.6\mu$m & Spitzer IRAC              & B. Tully              & 06-03-2013\\\hline
IC\,5052         & B         & ESO\,1m Schmidt            & ESO-LV project\\
                & H         & 3.9m Anglo-Australian Telescope & S. Ryder \& C. Tinney & 27-07-2002\\
                & I         & ANU 40inch Telescope      & J.C. O'Brien          & 10-07-2002\\
                & J         & UK 48-inch Schmidt        & Digitized Sky Survey  & 14-09-1976\\
                & R         & ESO\,1m Schmidt            & ESO-LV project\\
                & V         & ANU 40inch Telescope      & J.C. O'Brien          & 12-07-2002\\
                & $3.6\mu$m & Spitzer IRAC              & R. de Jong            & 18-09-2005\\
                & $4.5\mu$m & Spitzer IRAC              & R. de Jong            & 18-09-2005\\\hline
IC\,5249         & B         & ESO\,1m Schmidt            & ESO-LV project\\
                & H         & 3.9m Anglo-Australian Telescope & S. Ryder \& C. Tinney & 28-07-2002\\
                & I         & ANU 40inch Telescope      & J.C. O'Brien          & 15-07-2002\\
                & J         & UK 48-inch Schmidt        & Digitized Sky Survey  & 30-07-1978\\
                & Ks        & 3.9m Anglo-Australian Telescope & S. Ryder \& C. Tinney & 28-07-2002\\
                & R         & ANU 40inch Telescope      & J.C. O'Brien          & 11-07-2002\\
                & V         & ANU 40inch Telescope      & J.C. O'Brien          & 12-07-2002\\
                & $3.6\mu$m & Spitzer IRAC              & K. Sheth              & 31-07-2010\\
                & $4.5\mu$m & Spitzer IRAC              & K. Sheth              & 31-07-2010\\\hline
ESO\,115-G021    & I         & ANU 40inch Telescope      & J.C. O'Brien          & 10-07-2002\\
                & J         & UK 48-inch Schmidt        & Digitized Sky Survey  & 09-09-1975\\
                & R         & CTIO 0.9 meter telescope  & J. Funes              & 14-09-2001\\ 
                & V         & ANU 40inch Telescope      & J.C. O'Brien          & 12-07-2002\\
                & $3.6\mu$m & Spitzer IRAC              & R. Kennicutt          & 12-09-2007\\
                & $4.5\mu$m & Spitzer IRAC              & R. Kennicutt          & 12-09-2007\\\hline
ESO\,138-G014    & B         & ESO\,1m Schmidt            & ESO-LV project\\
                & H         & 2.3m Advanced Technology Telescope & J.C. O'Brien& 16-07-2002\\
                & I         & ANU 40inch Telescope      & J.C. O'Brien          & 10-07-2002\\
                & R         & ESO\,1m Schmidt            & ESO-LV project\\
                & V         & ANU 40inch Telescope      & J.C. O'Brien          & 12-07-2002\\
                & $3.6\mu$m & Spitzer IRAC              & B. Tully              & 09-05-2012\\\hline
ESO\,146-G014    & I         & ANU 40inch Telescope      & J.C. O'Brien          & 10-07-2002\\
                & Ks        & 3.9m Anglo-Australian Telescope & S. Ryder \& C. Tinney & 28-07-2002\\
                & $3.6\mu$m & Spitzer IRAC              & G. Rieke              & 31-10-2006\\\hline
ESO\,274-G001    & J         & UK 48-inch Schmidt        & Digitized Sky Survey  & 14-03-1975\\
                & R         & Danish 1.54m              & J. Rossa              & 01-08-2000\\
                & V         & ANU 40inch Telescope      & J.C. O'Brien          & 12-07-2002\\
                & $3.6\mu$m & Spitzer IRAC              & B. Tully              & 10-05-2012\\\hline
UGC\,7321        & I         & ANU 40inch Telescope      & J.C. O'Brien          & 04-03-2002\\
                & J         & Palomar 48-inch Schmidt   & Digitized Sky Survey  & 23-03-1990\\
                & R         & ANU 40inch Telescope      & J.C. O'Brien          & 11-07-2002\\
                & $3.6\mu$m & Spitzer IRAC              & L. Matthews           & 26-12-2005\\
                & $4.5\mu$m & Spitzer IRAC              & L. Matthews           & 26-12-2005\\\hline\hline
\end{tabular}
\caption{Details of stellar observations}\label{tbl:stellarobservations}
\end{table*}

\begin{table*}
\resizebox{\textwidth}{!}{
\centering
\begin{tabular}{l|c|c|c|cccc|c|c}
Galaxy	&	Band	&	Magnitude	&	$B/D$	&	$L_\textrm{d}$ [$\odot$]	&	$L_\textrm{b}$ [$\odot$]	&	$L_\textrm{tot}$  [$\odot$]	& $M_\textrm{HI}/L_\textrm{tot}$&	$\chi^2$	&	Source	\\	\hline	\hline
IC\,2531	&	R	&	12.5	&	0.01	&	3.49E+09	&	3.49E+07	&	3.52E+09	&	2.1	&	0.47	&	\citet{Doyle2005A}	\\		
	&	I	&	11.9	&	0.07	&	2.97E+09	&	1.99E+08	&	3.17E+09	&	2.3	&	0.18	&	\citet{Doyle2005A}	\\		
	&	Wise1	&		&		&		&		&		&		&	0.07	&		\\	\hline	
IC\,5052 (left)	&	B	&	11.6	&	0.64	&	2.05E+09	&	1.31E+09	&	3.36E+09	&	0.3	&	0.26	&	\citet{Doyle2005A}	\\		
	&	H	&	9.1	&	2.00	&	9.41E+08	&	1.88E+09	&	2.82E+09	&	0.3	&	0.20	&	\citet{Jarrett2003A}	\\		
	&	I	&	11.4	&	2.00	&	2.23E+08	&	4.46E+08	&	6.69E+08	&	1.3	&	0.14	&	\citet{Doyle2005A}	\\		
	&	J	&	9.7	&	2.00	&	7.70E+08	&	1.54E+09	&	2.31E+09	&	0.4	&	0.13	&	\citet{Jarrett2003A}	\\		
	&	R	&	11.7	&	2.00	&	2.79E+08	&	5.58E+08	&	8.37E+08	&	1.1	&	0.10	&	\citet{Doyle2005A}	\\		
	&	V	&	11.0	&	0.69	&	1.76E+09	&	1.22E+09	&	2.98E+09	&	0.3	&	0.14	&	\citet{Vaucouleurs1991b}	\\		
	&	Wise1	&		&		&		&		&		&		&	0.12	&		\\		
	&	Wise2	&		&		&		&		&		&		&	0.16	&		\\	\hline	
IC\,5052 (right)	&	B	&	11.6	&	1.49	&	6.77E+08	&	1.01E+09	&	1.69E+09	&	0.5	&	0.14	&	\citet{Doyle2005A}	\\		
	&	H	&	9.1	&	2.00	&	8.62E+08	&	1.72E+09	&	2.59E+09	&	0.3	&	0.17	&	\citet{Jarrett2003A}	\\		
	&	I	&	11.4	&	1.62	&	1.92E+08	&	3.10E+08	&	5.02E+08	&	1.8	&	0.07	&	\citet{Doyle2005A}	\\		
	&	J	&	9.7	&	2.00	&	6.74E+08	&	1.35E+09	&	2.02E+09	&	0.4	&	0.07	&	\citet{Jarrett2003A}	\\		
	&	R	&	11.7	&	0.93	&	3.46E+08	&	3.21E+08	&	6.67E+08	&	1.3	&	0.08	&	\citet{Doyle2005A}	\\		
	&	V	&	11.0	&	2.00	&	5.96E+08	&	1.19E+09	&	1.79E+09	&	0.5	&	0.06	&	\citet{Vaucouleurs1991b}	\\		
	&	Wise1	&		&		&		&		&		&		&	0.07	&		\\		
	&	Wise2	&		&		&		&		&		&		&	0.06	&		\\	\hline	
IC\,5249	&	B	&	13.9	&	0.26	&	2.33E+09	&	5.96E+08	&	2.93E+09	&	1.9	&	0.42	&	\citet{Doyle2005A}	\\		
	&	H	&	12.3	&	0.85	&	6.40E+08	&	5.42E+08	&	1.18E+09	&	4.7	&	0.18	&	\citet{2MASSB}	\\		
	&	I	&	13.7	&	0.46	&	5.26E+08	&	2.42E+08	&	7.68E+08	&	7.3	&	0.30	&	\citet{Doyle2005A}	\\		
	&	J	&	13.1	&	0.10	&	7.60E+08	&	7.60E+07	&	8.36E+08	&	6.7	&	0.11	&	\citet{2MASSB}	\\		
	&	K	&	12.6	&	0.86	&	5.67E+08	&	4.85E+08	&	1.05E+09	&	5.3	&	0.41	&	\citet{2MASSB}	\\		
	&	R	&	13.9	&	0.72	&	5.37E+08	&	3.86E+08	&	9.23E+08	&	6.1	&	0.11	&	\citet{Doyle2005A}	\\		
	&	V	&		&		&		&		&		&		&	0.06	&		\\		
	&	Wise1	&		&		&		&		&		&		&	0.15	&		\\		
	&	Wise2	&		&		&		&		&		&		&	0.19	&		\\	\hline	
ESO\,115-G021	&	I	&	13.1	&	2.00	&	1.21E+07	&	2.42E+07	&	3.63E+07	&	17.1	&	0.11	&	\citet{Doyle2005A}	\\		
	&	J	&	14.1	&	0.01	&	7.01E+06	&	7.01E+04	&	7.08E+06	&	87.6	&	0.13	&	\citet{Jarrett2003A}	\\		
	&	R	&	13.1	&	0.54	&	3.21E+07	&	1.75E+07	&	4.96E+07	&	12.5	&	0.08	&	\citet{Doyle2005A}	\\		
	&	V	&	12.6	&	0.58	&	8.89E+07	&	5.18E+07	&	1.41E+08	&	4.4	&	0.15	&	\citet{Vaucouleurs1988}	\\		
	&	Wise1	&	10.5	&	1.08	&	1.03E+08	&	1.11E+08	&	2.14E+08	&	2.9	&	0.19	&	\citet{Dale2009A}	\\		
	&	Wise2	&	10.4	&	1.23	&	1.05E+08	&	1.30E+08	&	2.35E+08	&	2.6	&	0.35	&	\citet{Dale2009A}	\\	\hline	
ESO\,138-G014	&	B	&	12.8	&	0.10	&	4.17E+09	&	4.04E+08	&	4.57E+09	&	0.6	&	0.32	&	\citet{Doyle2005A}	\\		
	&	H	&	10.6	&	0.30	&	1.98E+09	&	5.88E+08	&	2.57E+09	&	1.1	&	0.43	&	\citet{Jarrett2003A}	\\		
	&	I	&	12.4	&	2.00	&	3.81E+08	&	7.62E+08	&	1.14E+09	&	2.5	&	0.07	&	\citet{Doyle2005A}	\\		
	&	R	&	12.7	&	0.01	&	1.40E+09	&	1.40E+07	&	1.41E+09	&	2.1	&	0.27	&	\citet{Doyle2005A}	\\		
	&	V	&		&		&		&		&		&		&	0.08	&		\\		
	&	Wise1	&		&		&		&		&		&		&	0.14	&		\\	\hline	
ESO\,146-G014	&	I	&	13.9	&	0.39	&	1.78E+08	&	6.87E+07	&	2.47E+08	&	7.7	&	0.14	&	\citet{Doyle2005A}	\\		
	&	K	&		&	0.87	&		&		&		&		&	1.08	&		\\		
	&	V	&	14.6	&	1.98	&	1.23E+08	&	2.43E+08	&	3.66E+08	&	5.2	&	0.17	&	\citet{Zackrisson2006}	\\	\hline	
ESO\,274-G001	&	V	&	10.1	&	0.56	&	9.91E+08	&	5.55E+08	&	1.55E+09	&	0.2	&	0.04	&	\citet{Vaucouleurs1991b}	\\		
	&	J	&	8.7	&	0.55	&	9.01E+08	&	4.96E+08	&	1.40E+09	&	0.2	&	0.08	&	\citet{Jarrett2003A}	\\		
	&	R	&	10.9	&	0.73	&	2.42E+08	&	1.77E+08	&	4.19E+08	&	0.8	&	0.07	&	\citet{Doyle2005A}	\\		
	&	Wise1	&		&		&		&		&		&		&	0.13	&		\\	\hline	
UGC\,7321	&	R	&	12.7	&	1.04	&	2.92E+08	&	3.05E+08	&	5.97E+08	&	1.6	&	0.05	&	\citet{Taylor2005}	\\		
	&	I	&	12.4	&	1.08	&	2.00E+08	&	2.16E+08	&	4.16E+08	&	2.4	&	0.04	&	\cite{Makarova1999A}	\\		
	&	J	&	11.5	&	0.01	&	5.00E+08	&	5.00E+06	&	5.05E+08	&	1.9	&	0.09	&	\citet{Jarrett2003A}	\\		
	&	Wise1	&	10.3	&	2.00	&	2.51E+08	&	5.02E+08	&	7.53E+08	&	1.3	&	0.02	&	\citet{Dale2009A}	\\		
	&	Wise2	&	10.3	&	2.00	&	2.57E+08	&	5.14E+08	&	7.71E+08	&	1.3	&	0.05	&	\citet{Dale2009A}	\\	\hline	\hline
\end{tabular}
}
      \caption[Stellar fit results per band]{For each filter band, the $\chi^2$ quality of the fit is shown. If available, the literature brightness is shown together with its source. Where this is available, the total brightness of the galaxy in that band is calculated}\label{tbl:stellar-main2}
\end{table*}

\bsp

\label{lastpage}

\end{document}